\documentclass[twocolumn, pra, showpacs,superscriptaddress]{revtex4}
\usepackage{amssymb}
\usepackage{graphicx}
\usepackage{dcolumn}
\usepackage{bm}
\usepackage{amsmath}

\begin{document}

\title{Analytical ground state for the Jaynes-Cummings model with the
ultrastrong coupling}
\author{Yuanwei Zhang}
\affiliation{State Key Laboratory of Quantum Optics and Quantum Optics Devices, College
of Physics and Electronic Engineering, Shanxi University, Taiyuan 030006, P.
R. China}
\affiliation{Institute of Theoretical Physics, Shanxi University, Taiyuan 030006, P. R.
China}
\author{Gang Chen \footnote{%
Corresponding author: chengang971@163.com}} \affiliation{State Key
Laboratory of Quantum Optics and Quantum Optics Devices, College of
Physics and Electronic Engineering, Shanxi University, Taiyuan
030006, P. R. China} \affiliation{Department of Physics, Shaoxing
University, Shaoxing 312000, P. R. China}
\author{Lixian Yu}
\affiliation{Department of Physics, Shaoxing University, Shaoxing
312000, P. R. China}
\author{Qifeng Liang}
\affiliation{Department of Physics, Shaoxing University, Shaoxing
312000, P. R. China}
\author{J. -Q. Liang}
\affiliation{Institute of Theoretical Physics, Shanxi University, Taiyuan 030006, P. R.
China}
\author{Suotang Jia}
\affiliation{State Key Laboratory of Quantum Optics and Quantum Optics Devices, College
of Physics and Electronic Engineering, Shanxi University, Taiyuan 030006, P.
R. China}

\begin{abstract}
We present a generalized variational method to analytically obtain the
ground-state properties of the unsolvable Jaynes-Cummings model with the
ultrastrong coupling. An explicit expression for the ground-state energy,
which agrees well with the numerical simulation in a wide range of the
experimental parameters, is given. In particular, the introduced method can
successfully solve this Jaynes-Cummings model with the positive detuning
(the atomic resonant level is larger than the photon frequency), which can
not be treated in the adiabatical approximation and the generalized
rotating-wave approximation. Finally, we also demonstrate analytically how
to control the mean photon number by means of the current experimental
parameters including the photon frequency, the coupling strength, and
especially the atomic resonant level.
\end{abstract}

\maketitle

The Jaynes-Cummings model, which describes the important interaction between
the atom and the photon of a quantized electromagnetic field, is a
fundamental model in quantum optics and condensed-matter physics as well as
in quantum information science. In the optical cavity quantum
electrodynamics, the atom-photon coupling strength is far smaller than the
photon frequency. As a result, the system dynamics can be well governed by
the Jaynes-Cummings model with the rotating-wave approximation (RWA) \cite%
{Jaynes}. In the case of the RWA, its energy spectrum and wavefunctions can
be solved exactly \cite{Shore}. With the rapid development of fabricated
technique in solid-state systems, the Jaynes-Cummings model can be realized
in semiconducting dots \cite{Imamoglu,Gunter,Anappara,Todorov} and
superconducting Josephson junctions \cite%
{Blais,Wallraff,Tian,Hofheinz,LaHaye,Fedorov}. More importantly, recent
experiment has reported the existence of the ultrastrong coupling with the
ratio $0.12$ between the coupling strength and the microwave photon
frequency \cite{Niemczyk}. Moreover, this ratio maybe approach unit due to
the current efforts \cite{Bourassa,Peropadre}.

However, in this ultrastrong coupling regime the well-known RWA breaks down
and the whole Hamiltonian is written as
\begin{equation}
H_{x}=\omega (a^{\dagger }a+\frac{1}{2})+\frac{1}{2}\Omega \sigma
_{z}+g(\sigma _{+}+\sigma _{-})(a^{\dagger }+a),  \label{Without RWA}
\end{equation}%
where $a^{\dagger }$ and $a$\ are creation and annihilation operators for
photon with frequency $\omega $, $\sigma _{\pm }$ are the raising and
lowering operators of the two-level atom in the basis of $\sigma _{z}$, $%
\Omega $ is the atomic resonant frequency and $g$ is the atom-photon
coupling strength. Due to the existence of the counter-rotating terms ($%
\sigma _{+}a^{+}$ and $\sigma _{-}a$), Hamiltonian (\ref{Without RWA}) is
very difficult to be solved analytically except for $\Omega =0$. Although
the energy spectrum of Hamiltonian (\ref{Without RWA}) has been obtained
perfectly by means of the numerical simulation \cite{Feranchuk,Chen1,Chen2},
the analytical solutions are very necessary for extracting the fundamental
physics as well as in processing quantum information \cite%
{Zheng,Ashhab,Chen3}. In the negative detuning $\Omega <\omega $, the
adiabatic approximation method that the second term of Hamiltonian (\ref%
{Without RWA}) is treated as a small perturbation has been considered \cite%
{Irish1}. Recently, a generalized rotating-wave approximation (GRWA) has
also been proposed to solve Hamiltonian (\ref{Without RWA}) in the displaced
oscillator basis states \cite{Irish2}. This method can derive an analytical
energy of Hamiltonian (\ref{Without RWA}) in the ultrastrong coupling
successfully. However, it, like the adiabatic approximation method, is
invalid in the case of the positive detuning $(\Omega >\omega )$ \cite%
{Hausinger}. Moreover, the GRWA also leads to an unphysical relation that
the mean photon number is independent of the atomic resonant frequency.
Thus, it is very necessary to put forward a new method to re-consider
Hamiltonian (\ref{Without RWA}).

In this brief paper we present a generalized variational method (GVM) to
analytically obtain the ground-state properties of Hamiltonian (\ref{Without
RWA}) in the ultrastrong coupling regime \cite{Niemczyk}. An explicit
expression for the ground-state energy, which agrees well with the numerical
simulation in a wide range of the experimental parameters, is given. More
importantly, our method is valid for all regions of the atomic resonant
level including the negative detuning $(\Omega <\omega )$, the resonant case
$(\Omega =\omega )$, and especially the positive detuning $(\Omega >\omega )$%
. Finally, we also demonstrate analytically that the mean photon number is
strongly dependent on all parameters of Hamiltonian (\ref{Without RWA})
including the photon frequency, the coupling strength, and the atomic
resonant level. For a weak atomic resonant frequency, the mean photon number
depends linearly on it.

We first perform a rotation around the y-axis to rewrite Hamiltonian (\ref%
{Without RWA}) as $H_{z}=\omega (a^{\dagger }a+\frac{1}{2})+\frac{1}{2}%
\Omega \sigma _{x}-g\sigma _{z}(a^{\dag }+a)$. Under a unitary
transformation $U=\exp [\lambda \sigma _{z}(a^{\dag }-a)]$, the above
Hamiltonian $H_{z}$ becomes
\begin{equation}
H_{u}=\lambda ^{2}\omega +2\lambda g+H_{a}+H_{g}+H_{\Omega }  \label{HU}
\end{equation}%
with $H_{a}=\omega (a^{\dagger }a+\frac{1}{2})$, $H_{g}=-(g+\lambda \omega
)(a^{\dag }+a)\sigma _{z}$, and $H_{\Omega }=\frac{1}{2}\Omega \{\sigma
_{+}\exp [2\lambda (a^{\dag }-a)]+\sigma _{-}\exp [-2\lambda (a^{\dag
}-a)]\} $, where $\lambda $ is a dimensionless parameter. In the
Jaynes-Cummings model with the RWA, the dimensionless parameter $\lambda $
can be determined by setting $H_{g}=0$ and an effective Hamiltonian can be
obtained by series expansion with respect to $g/\omega $ \cite{Blais}.
However, in the ultrastrong coupling, this derivation is not valid.
Moreover, the terms $H_{g}$ and $H_{\Omega }$ play the important role in the
energy spectrum. As will be shown, the dimensionless parameter $\lambda $ in
the ultrastrong coupling can be determined analytically.

We now choose the basis states $\left\vert \pm ,N\right\rangle =\frac{1}{%
\sqrt{2}}\left( c_{e}^{\dagger }\pm c_{g}^{\dag }\right) \left\vert
0\right\rangle _{a}\left\vert N\right\rangle $, where $c_{e}^{\dagger }$ and
$c_{g}^{\dag }$ are the creation operators of the excited- and ground-
states, $\left\vert 0\right\rangle _{a}$ is the vacuum state of atom and $%
\left\vert N\right\rangle $ is the Fock state of the photon, to calculate
the energy spectrum by means of the perturbation method \cite{LF}. This
method requires the diagonal and non-diagonal matrix elements of Hamiltonian
(\ref{HU}) in the basis states. Since $-(g+\lambda \omega )\left\langle
N,\pm \right\vert (a^{\dag }+a)\sigma _{z}\left\vert N,\pm \right\rangle =0$
and $\omega \left\langle N,\pm \right\vert (a^{\dagger }a+\frac{1}{2}%
)\left\vert N,\pm \right\rangle =\omega (N+\frac{1}{2})$, the diagonal part
is given by
\begin{equation}
H_{0}=\lambda ^{2}\omega +2\lambda g+\omega (N+\frac{1}{2})+\left\langle
H_{\Omega }\right\rangle  \label{Hzero}
\end{equation}%
with $\left\langle H_{\Omega }\right\rangle =\left\langle N,\pm \right\vert
H_{\Omega }\left\vert N,\pm \right\rangle =\mp F(\lambda )L_{N}^{0}\left(
4\lambda ^{2}\right) $, where $F(\lambda )=-\frac{1}{2}\Omega \exp \left(
-2\lambda ^{2}\right) $ and $L_{i}^{j}(x)$ is the associated Laguerre
polynomials. The non-diagonal parts must be divided into two cases including
$\left\langle N,\pm \right\vert H_{g}\left\vert \pm ,M\right\rangle $ and $%
\left\langle N,\pm \right\vert H_{\Omega }\left\vert \pm ,M\right\rangle $
for even $(N-M)$, and $\left\langle N,\pm \right\vert H_{g}\left\vert \mp
,M\right\rangle $ and $\left\langle N,\pm \right\vert H_{\Omega }\left\vert
\mp ,M\right\rangle $ for odd $(N-M)$ since $\left\langle N,\pm \right\vert
H_{a}\left\vert \pm ,M\right\rangle =\left\langle N,\pm \right\vert
H_{a}\left\vert \mp ,M\right\rangle =0$. When $(N-M)$ is even, we have $%
\left\langle N,\pm \right\vert H_{g}\left\vert \pm ,M\right\rangle =0$ and $%
\left\langle N,\pm \right\vert H_{\Omega }\left\vert \pm ,M\right\rangle
=P_{N,M}$, where $P_{N,M}=\mp F(\lambda )\left( 2\lambda \right) ^{N-M}\sqrt{%
M!/N!}L_{M}^{N-M}\left( 4\lambda ^{2}\right) $. When $(N-M)$ is odd, the
matrix elements become $\left\langle N,\pm \right\vert H_{g}\left\vert \mp
,M\right\rangle =-\sqrt{N}\omega (g/\omega +\lambda )\delta _{N,M+1}$ and $%
\left\langle N,\pm \right\vert H_{\Omega }\left\vert \mp ,M\right\rangle
=-P_{N,M}$. From the above calculation we assume $N>M$ in terms of the
Hermitian of Hamiltonian (\ref{HU}).

By means of the matrix form of Hamiltonian (\ref{HU}) with the derived
diagonal and non-diagonal terms, the energy spectrum and wavefunction can,
in principle, be obtained. However, the procedure is very complicated.
Fortunately, in the ultrastrong coupling regime, the ground-state properties
can be calculated by the perturbation expansion of the non-diagonal terms.
In this method, the diagonal part is regarded as the unperturbed Hamiltonian
$H_{0}$ and the remaining part of Hamiltonian, $H_{r}=H_{u}-H_{0}$ is
treated as a perturbation \cite{LF}.

Formally, the ground-state energy can be written as \cite{Note}
\begin{equation}
E_{0}=E_{0}^{\left( 0\right) }+E_{0}^{\left( 1\right) }+E_{0}^{\left(
2\right) }.  \label{Energy}
\end{equation}%
Here $E_{0}^{\left( 0\right) }$ is the unperturbed ground-state energy.
According to Eq. (\ref{Hzero}), this unperturbed ground-state energy is
given by $E_{0}^{\left( 0\right) }=\frac{1}{2}\omega +\lambda ^{2}\omega
+2\lambda g+F(\lambda )$ for $N=0$. The first-order correction to the
unperturbed ground-state energy is given by $E_{0}^{\left( 1\right) }=0$
since $H_{r}$ has no diagonal matrix element. While the second-order
correction is evaluated as $E_{0}^{\left( 2\right) }=-\left[ -\left(
g+\omega \lambda \right) +2\lambda F(\lambda )\right] ^{2}/[\omega
-2F(\lambda )\left( 1-2\lambda ^{2}\right) ]-\sum_{N=2}^{\infty
}F^{2}(\lambda )\left( 2\lambda \right) ^{2N}/N![E_{e,N}^{\left( 0\right)
}-E_{0}^{\left( 0\right) }]$, where $e=\pm N$ is odd (even). On the other
hand, according to the spirit of the variational method \cite{LF}, the
dimensionless parameter $\lambda $ can be derived from minimizing the
unperturbed ground-state energy $E_{0}^{\left( 0\right) }$ by
\begin{equation}
\lambda \lbrack \omega +\Omega \exp (-2\lambda ^{2})]+g=0.  \label{Planta}
\end{equation}

\begin{figure}[tp]
\includegraphics[width=8cm]{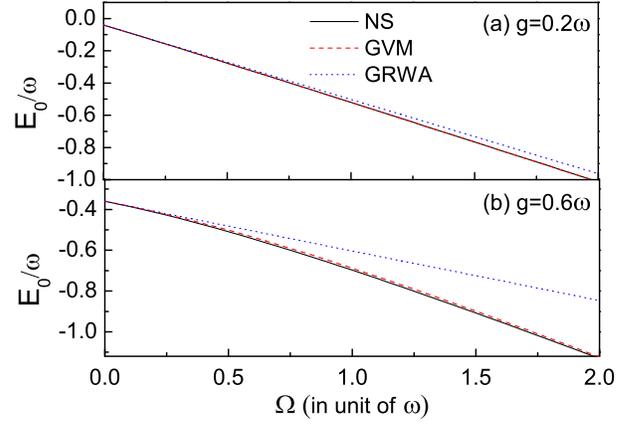}\newline
\caption{(Color online) The ground-state energy $E_{0}$ as a function of the
atomic resonant frequency $\Omega $ for the different coupling strength $%
g=0.2\protect\omega$ (a) and $0.6\protect\omega$ (b) by means of the
numerical simulation (NS), the GVM and the GRWA.}
\label{fig1}
\end{figure}

By using the same procedure, the ground-state wavefunction is written by
\cite{Note}
\begin{equation}
\left\vert \Psi _{0}\right\rangle =\left\vert \Psi _{0}^{\left( 0\right)
}\right\rangle +\left\vert \Psi _{0}^{\left( 1\right) }\right\rangle ,
\label{Wavef}
\end{equation}%
where the unperturbed ground-state wavefunction is given by $\left\vert \Psi
_{0}^{\left( 0\right) }\right\rangle =\left\vert -,0\right\rangle $, and the
first-order correction to the unperturbed ground-state wavefunction is given
by $\left\vert \Psi _{0}^{\left( 1\right) }\right\rangle =-\frac{\left[
-\left( g+\omega \lambda \right) +2\lambda F(\lambda )\right] }{\omega
-2F(\lambda )\left( 1-2\lambda ^{2}\right) }\left\vert +,1\right\rangle
-\sum_{N=2,4,\cdots }^{\infty }\frac{F(\lambda )\left( 2\lambda \right) ^{N}%
}{\sqrt{N!}[E_{-,N}^{\left( 0\right) }-E_{0}^{\left( 0\right) }]}\left\vert
-,N\right\rangle -\sum_{N=3,5,\cdots }^{\infty }\frac{F(\lambda )\left(
2\lambda \right) ^{N}}{\sqrt{N!}[E_{+,N}^{\left( 0\right) }-E_{0}^{\left(
0\right) }]}\left\vert +,N\right\rangle $.

Equations (\ref{Energy}) and (\ref{Wavef}), together with Eq. (\ref{Planta}%
), are the main results in this brief paper. Interestingly, the
dimensionless parameter $\lambda $ depends on all parameters including the
photon frequency $\omega $, the coupling strength $g$, and especially the
atomic resonant frequency $\Omega $. In general, Eq. (\ref{Planta}) can not
be solved and the ground-state properties can not be obtained analytically
due to the existence the complex sums in the expressions $E_{0}^{\left(
2\right) }$ and $\left\vert \Psi _{0}^{\left( 1\right) }\right\rangle $.
However, in current experimental setups of the ultrastrong coupling (for
example, $g\simeq 0.1\omega $ in Ref. \cite{Niemczyk}), Eq. (\ref{Planta})
becomes very simple, and moreover, the complex sums are also simplified
successfully. As a consequence, some interesting results, which govern the
fundamental properties of Hamiltonian (\ref{Without RWA}), can be achieved.

\begin{figure}[tp]
\includegraphics[width=8cm]{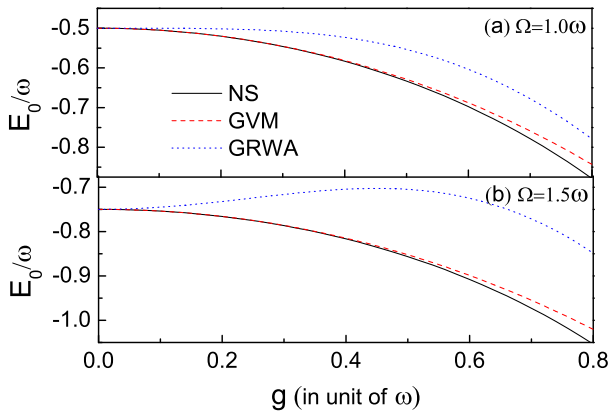}\newline
\caption{(Color online) The ground-state energy $E_{0}$ as a function of the
coupling strength $g$ for the different atomic resonant frequency $\Omega
=1.0\protect\omega$ (a), and $1.5\protect\omega$ (b) by means of the
numerical simulation (NS), the GVM, and the GRWA.}
\label{fig2}
\end{figure}

By means of Eq. (\ref{Planta}), the dimensionless parameter $\lambda $ is
derived approximately by $\lambda \simeq -g/\{\omega +\Omega \exp
[-2g^{2}/(\omega +\Omega )^{2}]\}$. However, in current experimental setups
of the ultrastrong coupling, $\exp [-2g^{2}/(\omega +\Omega )^{2}]\simeq 1$,
and thus, the dimensionless parameter $\lambda $ becomes
\begin{equation}
\lambda =-\frac{g}{\omega +\Omega }.  \label{POB}
\end{equation}%
Based on Eq. (\ref{POB}), we have $E_{0}^{\left( 0\right) }=\omega
/2-g^{2}\left( \omega +2\Omega \right) /\left( \omega +\Omega \right)
^{2}-\Omega \exp [-2g^{2}/\left( \omega +\Omega \right) ^{2}]/2$ and $%
E_{0}^{\left( 2\right) }\approx -4g^{2}\Omega ^{2}\lambda ^{4}/\left( \omega
+\Omega \right) ^{3}-\sum_{N=2}^{\infty }F^{2}(\lambda )\left( 2\lambda
\right) ^{2N}/N![E_{e,N}^{\left( 0\right) }-E_{-,0}^{\left( 0\right) }]$.
Due to the existence of the high order terms with respect to $\lambda $, the
complex sum in $E_{0}^{\left( 2\right) }$ can be omitted. Moreover, the
first term of $E_{0}^{\left( 2\right) }$ is far smaller than $E_{0}^{\left(
0\right) }$. It means that the second-order perturbed energy almost has no
effect to the ground-state energy. Therefore, an explicit ground-state
energy for Hamiltonian (\ref{Without RWA}) is given finally by
\begin{equation}
E_{0}\simeq \frac{\omega }{2}-\frac{g^{2}\left( \omega +2\Omega \right) }{%
\left( \omega +\Omega \right) ^{2}}-\frac{\Omega }{2}\exp [-2(\frac{g}{%
\omega +\Omega })^{2}].  \label{GN}
\end{equation}%
For a weak atomic resonant frequency $(\Omega \ll \omega )$, Eq. (\ref{GN})
become $E_{0}^{GRWA}=1/2\omega -g^{2}/\omega -1/2\Omega \exp [-2\left(
g/\omega \right) ^{2}]$, which has been obtained perfectly by means of the
GRWA \cite{Irish2}.

\begin{figure}[tp]
\includegraphics[width=7.3cm]{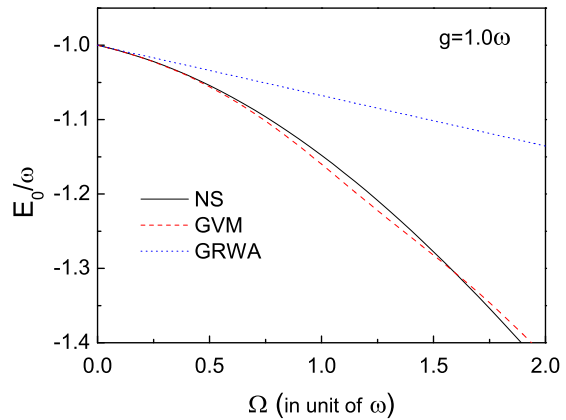}\newline
\caption{(Color online) The ground-state energy $E_{0}$ as a function of the
atomic resonant level $\Omega $ with the coupling strength $g=1.0\protect%
\omega $ by means of the numerical simulation (NS), the GVM, and the GRWA. }
\label{fig3}
\end{figure}

Figure 1 shows the ground-state energy $E_{0}$ as a function of the atomic
resonant frequency $\Omega $ for the different coupling strength $%
g=0.2\omega $ and $0.6\omega$ by means of the numerical simulation, the GVM (%
$E_{0}$) and the GRWA ($E_{0}^{GRWA}$). For a weak atomic resonant frequency
$(\Omega <$ $\omega )$, both results derived from the GVM and the GRWA agree
with the direct numerical simulation. In the positive detuning ($\Omega
>\omega $), the GRWA breaks down and the error is increased linearly.
However, the result derived from the GVM can also agree well with that of
the numerical simulation. Figure 2 shows the ground-state energy $E_{0}$ as
a function of the coupling strength $g$ for different atomic resonant
frequency $\Omega =1.0\omega$ and $1.5\omega$ by means of the numerical
simulation, the GVM ($E_{0}$) and the GRWA ($E_{0}^{GRWA}$). This figure
exhibits that the analytical expression in Eq. (\ref{GN}) can work
reasonably for $g<0.8\omega $. If $g\sim \omega $, Eq. (\ref{Planta}) can
not be solved explicitly and the complex sums in the expressions $%
E_{0}^{\left( 2\right) }$ and $\left\vert \Psi _{0}^{\left( 1\right)
}\right\rangle $ can also not be simplified. However, the results in Eqs. (%
\ref{Energy}) and (\ref{Wavef}) with the complex sums remain valid, as shown
in Fig. 3.

Having obtained the ground-state energy, we now discuss the
experimentally-measurable mean photon number $\left\langle a^{\dag
}a\right\rangle $, which can be obtained by $\left\langle a^{\dag
}a\right\rangle =\left\langle \Psi _{0}\right\vert Ua^{\dag
}aU^{\dag }\left\vert \Psi _{0}\right\rangle =\left\langle \Psi
_{0}\right\vert a^{\dag }a+\lambda ^{2}-\lambda \sigma _{z}\left(
a^{\dag }+a\right) \left\vert \Psi _{0}\right\rangle $. The
expression of the mean photon number is very complex. However, here
we give an approximate solution,
\begin{equation}
\left\langle a^{\dag }a\right\rangle \simeq \frac{g^{2}}{[\omega +\Omega
\exp (-2g^{2}/\omega ^{2})]^{2}}.  \label{MAP}
\end{equation}%
Equation (\ref{MAP}) demonstrates clearly that the mean photon number
depends strongly on all parameters of Hamiltonian (\ref{Without RWA}). For a
weak atomic resonant frequency $(\Omega \ll \omega )$, the mean photon
number becomes $\left\langle a^{\dag }a\right\rangle =g^{2}/\omega
^{2}-2g^{2}\Omega \exp (-2g^{2}/\omega ^{2})/\omega ^{3}$, which is quite
different from the previous conclusion $\left\langle a^{\dag }a\right\rangle
_{GRWA}=g^{2}/\omega ^{2}$ derived from the GRWA \cite{Irish2}.

Figures 4 shows the mean photon number $\left\langle a^{\dag }a\right\rangle
$ as a function of the atomic resonant frequency $\Omega $ for the coupling
strength $g=0.6\omega $ by means of the numerical simulation, the GVM and
the GRWA. In the insert of Fig. 4, the mean photon number $\left\langle
a^{\dag }a\right\rangle $ as a function of the coupling strength $g$ is also
plotted when the atomic resonant frequency is chosen as $\Omega =1.5\omega $%
. Both these figures show that our obtained explicit expression in Eq.(\ref%
{MAP}) agrees well with the numerical simulation. It implies that the
required mean photon number can controlled well by manipulating an external
flux bias (the atomic resonant frequency), the inductive coupling of a qubit
(the coupling strength), and a transmission line resonator (the photon
frequency) in current experimental setups of the ultrastrong coupling \cite%
{Niemczyk}.

\begin{figure}[tp]
\includegraphics[width=7cm]{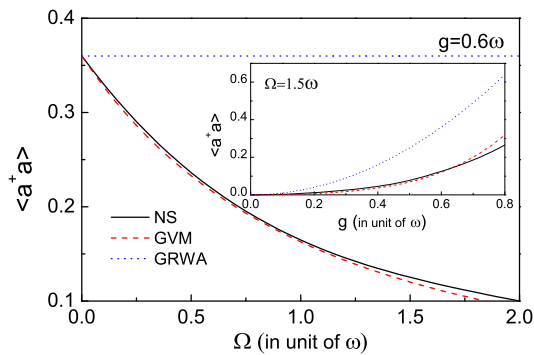}\newline
\caption{(Color online) The mean photon number $\left\langle a^{\dag
}a\right\rangle $ as a function of the atomic resonant frequency $\Omega $
for the coupling strength $g=0.6\protect\omega$ by means of the numerical
simulation (NS), the GVM, and the GRWA. Insert: The mean photon number $%
\left\langle a^{\dag }a\right\rangle $ as a function of the coupling
strength $g$ for the atomic resonant frequency $\Omega=1.5\protect\omega$.}
\label{fig4}
\end{figure}

In summary, we have presented the GVM to analytically obtain some
interesting ground-state properties of the Jaynes-Cummings model with the
ultrastrong coupling in all regions of the atomic resonant level including
the negative detuning, the resonant case, and especially the positive
detuning. However, for the excited state, this method is invalid since the
second-order perturbation energy will be larger than the unperturbed
ground-state energy.

We thank Drs. Chuanwei Zhang, Ming Gong, and Yongping Zhang for helpful
discussions and suggestions. This work was supported partly by the NNSFC
under Grant Nos. 10904092, 10934004, 60978018, 11074154, and 11075099, and
the ZJNSF under Grant No. Y6090001.


\begin{thebibliography}{99}
\bibitem{Jaynes} E. T. Jaynes, and F. W. Cummings, Proc. IEEE \textbf{51},
89 (1963).

\bibitem{Shore} B. W. Shore, and P. L. Knight, J. Mod. Opt. \textbf{40},
1195 (1993).

\bibitem{Imamoglu} A. Imamoglu, D. D. Awschalom, G. Burkard, D. P.
DiVincenzo, D. Loss, M. Sherwin, and A. Small, Phys. Rev. Lett. \textbf{83},
4204 (1999).

\bibitem{Gunter} G. G\"{u}nter, A. A. Anappara, J. Hees, A. Sell1, G.
Biasiol, L. Sorba, S. De Liberato, C. Ciuti, A. Tredicucci, A.
Leitenstorfer, and R. Huber, Nature (London) \textbf{458}, 178 (2009).

\bibitem{Anappara} A. A. Anappara, S. D. Liberato, A. Tredicucci, C. Ciuti,
G. Biasio, L. Sorba, and F. Beltram, Phys. Rev. B \textbf{79}, 201303, (2009)

\bibitem{Todorov} Y. Todorov, A. M. Andrews, R. Colombelli, S. De Liberato,
C. Ciuti, P. Klang, G. Strasser, and C. Sirtori, Phys. Rev. Lett. \textbf{105%
}, 196402 (2010).

\bibitem{Blais} A. Blais, R.-S. Huang, A. Wallraff, S. M. Girvin, and R. J.
Schoelkopf, Phys. Rev. A \textbf{69}, 062320 (2004).

\bibitem{Wallraff} A. Wallraff, D. I. Schuster, A. Blais, L. Frunzio, R.- S.
Huang, J. Majer, S. Kumar, S. M. Girvin, and R. J. Schoelkopf, Nature
(London) \textbf{431}, 162 (2004).

\bibitem{Tian} L. Tian, P. Rabl, R. Blatt, and P. Zoller, Phys. Rev. Lett.
\textbf{92}, 247902 (2004).

\bibitem{Hofheinz} M. Hofheinz, H. Wang, M. Ansmann, R. C. Bialczak, E.
Lucero, M. Neeley, A. D. O'Connell, D. Sank, J. Wenner, John M. Martinis,
and A. N. Cleland, Nature (London) \textbf{459}, 546 (2009).

\bibitem{LaHaye} M. D. LaHaye, J. Suh, P. M. Echternach, K. C. Schwab, and
M. L. Roukes, Nature (London) \textbf{459}, 960(2009).

\bibitem{Fedorov} A. Fedorov, A. K. Feofanov, P. Macha, P. Forn-D\'{\i}az,
C. J. P. M. Harmans, and J. E. Mooij, Phys. Rev. Lett. \textbf{105}, 060503
(2010).

\bibitem{Niemczyk} T. Niemczyk, F. Deppe, H. Huebl, E. P. Menzel, F. Hocke,
M. J. Schwarz, J. J. Garcia-Ripoll, D. Zueco, T. H\"{u}mmer, E. Solano, A.
Marx, and R. Gross, Nature Physics \textbf{6}, 772 (2010).

\bibitem{Bourassa} J. Bourassa, J. M. Gambetta, A. A. Abdumalikov, Jr., O.
Astafiev,Y. Nakamura, and A. Blais, Phys. Rev. A. \textbf{80}, 032109 (2009).

\bibitem{Peropadre} B. Peropadre, P. Forn-D\'{\i}az, E. Solano, and J. J.
Garc\'{\i}a-Ripoll, Phys. Rev. Lett. \textbf{105}, 023601 (2010).

\bibitem{Feranchuk} I. D. Feranchuk, L. I. Komarov, and A. P. Ulyanenkov, J.
Phys. A: Math. Gen. \textbf{29}, 4035 (1996).

\bibitem{Chen1} Q. H. Chen, L. Li, T. Liu, and K. -L. Wang, arXiv: 1007.1747.

\bibitem{Chen2} Q. H. Chen, T. Liu, Y. -Y. Zhang, and K. -L. Wang, arXiv:
1011.3280.

\bibitem{Zheng} H. Zheng, S. Y. Zhu, and M. S. Zubairy, Phys. Rev. Lett.
\textbf{101}, 200404 (2008).

\bibitem{Ashhab} S. Ashhab, and F. Nori, Phys.\ Rev. A \textbf{81}, 042311
(2010).

\bibitem{Chen3} Q. H. Chen,Y. Yang, T. Liu, and K. -L. Wang, Phys.\ Rev. A
\textbf{82}, 052306 (2010).

\bibitem{Irish1} E. K. Irish, J. Gea-Banacloche, I. Martin, and K. C. Schwa,
Phys. Rev. B. \textbf{72,} 195410 (2005).

\bibitem{Irish2} E. K. Irish, Phys. Rev. Lett. \textbf{99}, 173601 (2007).

\bibitem{Hausinger} J. Hausinger, and M. Grifoni, Phys. Rev. A \textbf{82},
062320 (2010).

\bibitem{LF} I. G. Lang, and Yu. A. Firsov, Zh. Eksp. Teor. Fiz. \textbf{43}%
,1843 (1962) [Sov. Phys. JETP \textbf{16}, 1301 (1963)].

\bibitem{Note} In this brief paper the ground-state energy is only
considered up to the second order and the ground-state wavefunction is only
considered up to the first order. These results can agree well with those of
the numerical simulations. The detailed comparison is shown in Figs. (1)-(4).
\end{thebibliography}
\end{document}